\documentclass[twocolumn,prl,preprintnumbers,amsmath,amssymb]{revtex4}

\usepackage{graphicx}% Include figure files
\usepackage{dcolumn}% Align table columns on decimal point
\usepackage{bm}% bold math
\usepackage{subfigure}
\usepackage{float}
\usepackage{amsfonts}
\usepackage{amsmath}
\usepackage{amssymb,epsf}
\usepackage{latexsym}
\usepackage{graphicx,epsfig}
\usepackage{amssymb}
\usepackage{subfigure}
\usepackage[colorlinks=true,citecolor=blue,linkcolor=blue,urlcolor=black]{hyperref}
\usepackage{epstopdf}
\usepackage{color}

\begin{document}
%\preprint{APS/123-QED}
\title{\textbf{\textrm{Optimal reinforcement learning near the edge of synchronization transition}}}
\author{Mahsa Khoshkhou}
\author{Afshin Montakhab}
\email{montakhab@shirazu.ac.ir}
\affiliation{Department of Physics, College of Sciences, Shiraz University, Shiraz 71946-84795, Iran}
\date{\today}
%%%-----------------------------------------------------------------------------
\begin{abstract}
Recent experimental and theoretical studies have indicated that
the putative criticality of cortical dynamics may corresponds to a
synchronization phase transition. The critical dynamics near such
a critical point needs further investigation specifically when
compared to the critical behavior near the standard absorbing
state phase transition. Since the phenomena of learning and
self-organized criticality (SOC) at the edge of synchronization
transition can emerge jointly in spiking neural networks due to
the presence of spike-timing dependent plasticity (STDP), it is
tempting to ask: What is the relationship between synchronization
and learning in neural networks? Further, does learning benefit
from SOC at the edge of synchronization transition? In this paper,
we intend to address these important issues. Accordingly, we
construct a biologically inspired model of a cognitive system
which learns to perform stimulus-response tasks. We train this
system using a reinforcement learning rule implemented through
dopamine-modulated STDP. We find that the system exhibits a
continuous transition from synchronous to asynchronous neural
oscillations upon increasing the average axonal time delay. We
characterize the learning performance of the system and observe
that it is optimized near the synchronization transition.  We also
study neuronal avalanches in the system and provide evidence that
optimized learning is achieved in a slightly supercritical state.
\end{abstract}
%%%-----------------------------------------------------------------------------

%%%%%PACS means Physics and Astronomy Classification Scheme
%\keywords{}%Use showkeys class option if keyword
\maketitle
%--------------------------------------------------------------------------------

\section{Introduction}
A cognitive system, either biological or artificial, is a
continuously active complex system autonomously exploring and
reacting to the environment with the capability to survive
\cite{Gros2015}. In contrast with the usual paradigm of artificial
intelligence (AI) which mainly follows an all-in-one-step approach
to intelligent systems \cite{Russell2002,Jordan2015,Mohri2018}, a
cognitive system is not necessarily intelligent, but intelligence
can be achieved once the system has been developed
\cite{Gros2009}. The universal principles necessary for the
realization of a cognitive system resembling our own cognitive
organ is not fully understood, yet. However, it is believed that
learning within a \textcolor[rgb]{0.00,0.00,1.00}{biological}
cognitive system with self-induced dynamics is not supervised
\cite{Slijepcevic2021}. Accordingly, reinforcement learning (RL),
i.e. learning how to map situations to actions so as to maximize a
numerical reward signal, is the closest paradigm to the kind of
learning that \textcolor[rgb]{0.00,0.00,1.00}{biological} systems
follow \cite{SuttonBarto2018,Neffci2019}. Trial-and-error search
and delayed reward are the most important distinguishing
characteristics of RL \cite{SuttonBarto2018}.

Spike-timing dependent plasticity (STDP) is an experimentally
well-established rule that leads to synapse strengthening for
correlated activity at the pre- and postsynaptic neurons and
synapse weakening for decorrelated activity
\cite{Bi2001,Caporale2008,Markram2012,Brzosko2019,Dickey2021}. A modified version of this rule that is
called dopamine-modulated STDP is the main candidate to explain
the relationship between synaptic plasticity on the microscopic
level, and the adaptive changes of behavior of biological
organisms on the macroscopic level. i.e., linkage of dopamine
signaling with STDP triggered the development of phenomenological
models for RL that could explain how behaviorally relevant
adaptive changes in complex networks of spiking neurons could be
achieved in a self-organizing manner
\cite{IzhikevichDA,Legenstein2008a,Florian2007,Ferries2007,Gireesh2008}.

STDP is also identified as a candidate mechanism underlying the
emergence and maintenance of self-organized criticality (SOC) in
neural circuits \cite{Rubinov2011,Meisel2009,Kossio2018}. It is
hypothesized that cortical neural circuits self-organize to a
critical state which is associated with a transition point of a
continuous phase transition
\cite{Beggs2003,Beggs2004,Plenz2014,Legenstein2007}. Operating at
the vicinity of this critical point has functional benefits for a neural
system including optimal capacity of information processing,
transfer and storage as well as optimal dynamic range
\cite{Legenstein2007,Shew2013,Boedecker2012,Kinouchi2006}.
However, the relationship between SOC and learning is not clear.
Criticality has been shown to be beneficial for learning in some
models \cite{deArcangelis2010,Berger2017,vanKessenich2019}, but
destructive to learning \cite{DelPapa2017}, or having
task-dependent profit \cite{Cramer2020}, in other models.

As opposed to the earlier models which have attributed the
putative criticality of cortical dynamics to an activity
transition point associated with an absorbing state phase
transition
\cite{Levina2007,deArcangelis2006,Millman2010,Stepp2015}, recent
experimental and theoretical studies indicate that such critical
dynamics corresponds to  a synchronization phase transition, at
which incipient oscillations and scale-free avalanches coexist
\cite{PRL2019,KM2019,di Santo2018}. In particular, we have
previously shown that neurophysiological regulatory mechanisms
such as STDP with suitable axonal conduction delays brings about
and maintains SOC at the edge of synchronization transition in a
biologically meaningful model of cortical networks \cite{KM2019}.
The basic idea is that the positive feedback, where
synchronization leads to synaptic potentiation which in turn leads
to more synchronization, is broken with a time-delay.
Consequently, synaptic strengths depress and potentiate in turn as
the system finds and maintains a critical state at the edge of
synchronization with scale-free avalanches. The implications for
brain's performance at such a critical point has been investigated
in recent studies \cite{PRL2019,KM2019,di
Santo2018,Poil2012,DallaPorta2019}.

We note that the phenomena of
learning and SOC at the edge of synchronization are both rooted in
the regulatory mechanism of STDP. Therefore, it is natural to ask
what the relation between learning and edge of synchronization is.
In particular, can learning profit from brain operating at or near
a synchronization transition? This is exactly what we propose to
address in this article.

In order to specify the relationship between synchronization,
learning and SOC in cognitive systems, we construct a
biologically relevant model of cortical networks with parameters
that are set according to empirical evidence. Constructing a
biologically inspired model of a cognitive system is
of crucial importance because it is an attempt toward building an
artificial brain on one hand, and toward understanding the
relevance of various biological mechanisms in the brain on the
other hand. This cognitive system learns to perform
stimulus-response (SR) tasks with different levels of
complication. We train this system using a RL rule which is
implemented through dopamine-modulated STDP. We show that the
system exhibits a continuous transition from synchronous to
asynchronous neural oscillations upon increasing axonal time
delays. The learning performance of the system depends on the
amount of synchronization in neural activity. For all SR tasks of
various complexity, optimal performance is achieved while the
system is above but very close to the transition point of
synchronization. More strictly, the optimal performance occurs at
a slightly supercritical state of synchronization.

In the following section, we describe the model we use for our
study. Results of our numerical study is presented in section III,
and we close the paper with some concluding remarks in section IV.

\section{Model and Methods}
We construct a biologically inspired cognitive system.
A detail description of this system is given below.

\textbf{Neural dynamics-} The system consists of $N$ spiking
Izhikevich neurons interacting by transition of chemical synaptic
currents with axonal conduction delays. The dynamics of each
neuron is described by a set of two differential equations
\cite{Izhikevich2003}:
\begin{equation}\label{sequ1}
\dot{v}_i=0.04v_i^2+5v_i+140-u_i+I_i^{DC}+I_i^{syn}
\end{equation}
\begin{equation}\label{sequ2}
\dot{u}_i=a(bv_i-u_i)
\end{equation}
\\with the auxiliary after-spike reset:
\begin{equation}\label{sequ3}
\text{if}\ \ v_i{\geq}30,\ \ \text{then} \ v_i \ {\rightarrow} \ c \ \ \text{and}\ u_i \ {\rightarrow} \ u_i+d
\end{equation}
\\
for $i=1, 2,..., N$. Here, $v_i$ is the membrane potential and
$u_i$ is the membrane recovery variable. When $v_i$ reaches its
apex ($v_{max}=30$ mV), $v_i$ and $u_i$ are reset according to
Eq.(3). $a$, $b$, $c$ and $d$ are adjustable parameters that
determine the pattern of firing and are different for excitatory
and inhibitory neurons (see Table 1) \cite{Izhikevich2003}. The
population density of inhibitory neurons is set to be $\rho=0.2$.

\begin{table}[!b]
\caption{\small Values of constant parameters used in this study.}
\label{table1}
\begin{center}
\begin{tabular}{|m{1.6cm}|m{1.5cm} m{1.5cm}  m{1.5cm} m{1.6cm}|}
\hline
Izhikevich & $a^{ex}=0.02$ & $b^{ex}=0.2$ & $c^{ex}=-65$ & $d^{ex}=8$\\
neuron & $a^{in}=0.1$ & $b^{in}=0.2$ & $c^{in}=-65$ & $d^{in}=2$ \\
\hline
Synaptic current & $\tau_f=1$ &  $\tau_s=5$ & $V_0^{ex}=0$ & $V_0^{in}=-75$  \\
\hline
plasticity & $A_\pm=0.05$ &  $\tau_\pm=30$ & $w_{min}=0$ & $w_{max}=1$ \\
rule & $\tau_x$=1000 & $\tau_y=200$ & $y_0=2$ & \\
 \hline
\end{tabular}
\end{center}
\end{table}

The term $I_i^{DC}$ is an external current which determines
intrinsic firing rate of uncoupled neurons
\cite{Izhikevich2003,KM2018}. The values of $I_i^{DC}$ are chosen
randomly from the range $[3.8,4.5]$. This choice leads to
alpha-band intrinsic firing frequencies with a mean firing rate
around $9$ Hz. Alpha rhythms are typically observed during
learning and task performance
\cite{Bays2015,deGraaf2013,Grabner2004,Foxe2011}. The term
$I_i^{syn}$ represents the chemical synaptic current that goes
into each post-synaptic neuron $i$ \cite{Roth}:
\begin{equation}\label{sequ4}
I_i^{syn}=\frac{V_0-v_i}{D_i}\sum_jw_{ji}\frac{exp(-\frac{t-(t_j+\tau_{ji})}{\tau_s})-exp(-\frac{t-(t_j+\tau_{ji})}{\tau_f})}{\tau_s-\tau_f}
\end{equation}
\\
Here, $D_i$ is the in-degree of node $i$, $t_j$ is the instance of
last spike of pre-synaptic neuron $j$, and $\tau_{ji}$ is the
axonal conduction delay from pre-synaptic neuron $j$ to
post-synaptic neuron $i$. The values of $\tau_{ji}$ are chosen
randomly from a Poisson distribution with mean value $\tau=\langle
\tau_{ji} \rangle$. $\tau=0$ means that $\tau_{ji}=0$ $\forall$ $
j\neq i$.  $\tau_f$ and $\tau_s$ are the synaptic fast and slow
time constants and $V_0$ is the reversal potential of the synapse.
The most relevant variable in the context of learning is  $w_{ji}$
which denotes the strength of synapse from neuron $j$ to neuron
$i$. $w_{ji}$'s are the elements of the adjacency matrix of the
underlying network, ($w_{ji} \neq 0$ if there is a directed edge
from neuron $j$ to neuron $i$ and $w_{ji}=0$ otherwise). The
simulations are carried out using
Erdos-Renyi (ER) random networks
of size $N$ and average
connectivity $z=pN$ where we set the connection probability
$p=0.1$ \cite{Erdos1959}. The initial strength of excitatory
synapses is $w_{ji}(t=0)=w_0$. To balance the
excitation/inhibition in the system initial strength of inhibitory
synapses is set to $w_{ji}=r w_0$, while $r$ is the ratio of
excitatory to inhibitory synapses.

\textbf{Dopamine-modulated STDP rule-} To train the network, we
modify the strength of \emph{excitatory} synapses using the
simplest phenomenological model that captures the essence of
dopamine modulation of STDP \cite{IzhikevichDA}. According to this
model:
\begin{equation}
\dot{w}_{ji}= x_{ji} y
\end{equation}
\begin{equation}
\dot{x}_{ji}= -\frac{x_{ji}}{ \tau_x} + \Gamma_{ji}(\Delta t) \delta(t- t_{pre/post})
\end{equation}
\begin{equation}
\dot{y}= -\frac{y}{\tau_y} + y_0 \delta (t-t_{rew})
\end{equation}
\begin{eqnarray}
\Gamma_{ji}(\Delta t) = \left \{
\begin{array}{cc}
A_+(w_{max}-w_{ji})e^{-\frac{\Delta t-\tau_{ji}}{\tau_+}} & \text{if} \ \Delta t>\tau_{ji}  \\
-A_-(w_{ji}-w_{min})e^{\frac{\Delta t-\tau_{ji}}{\tau_-}} & \text{if} \ \Delta t \leq \tau_{ji}
\end{array}
\right.
\end{eqnarray}
\\
Here, $x_{ji}$ is the synaptic eligibility function or synaptic
tag, $y$ is the extracellular concentration of dopamine which is
assumed to be the same for all synapses, $\tau_x$ and $\tau_y$ are
time constants and $\delta(.)$ is the Dirac delta function.
$t_{rew}$ is the time when the reward is delivered to the network,
and $y_0$ is the amount of dopamine released by  the activity of
dopaminergic neurons at $t=t_{rew}$.  $\Gamma(\Delta t)$ is a
nearest-neighbor and soft-band STDP function with upper and lower
bands $w_{max}$ and $w_{min}$ \cite{Markram2012,Bi2001}. $\Delta
t=t_{post}-t_{pre}$ is the time difference between the last post-
and pre-synaptic spikes, $A_{\pm}$ determine the maximum synaptic
potentiation and depression, and $\tau_{\pm}$ determine the
temporal extent of the STDP window for potentiation and
depression. Numerical values of all of these parameters are listed
in Table I. Fig.1 illustrates how dopamine-modulated STDP leads to
potentiation of a synapse that plays a role in the correct
network's response.

\begin{figure}[!tb]
\begin{center}
{\includegraphics[width=0.50\textwidth,height=0.25\textwidth]{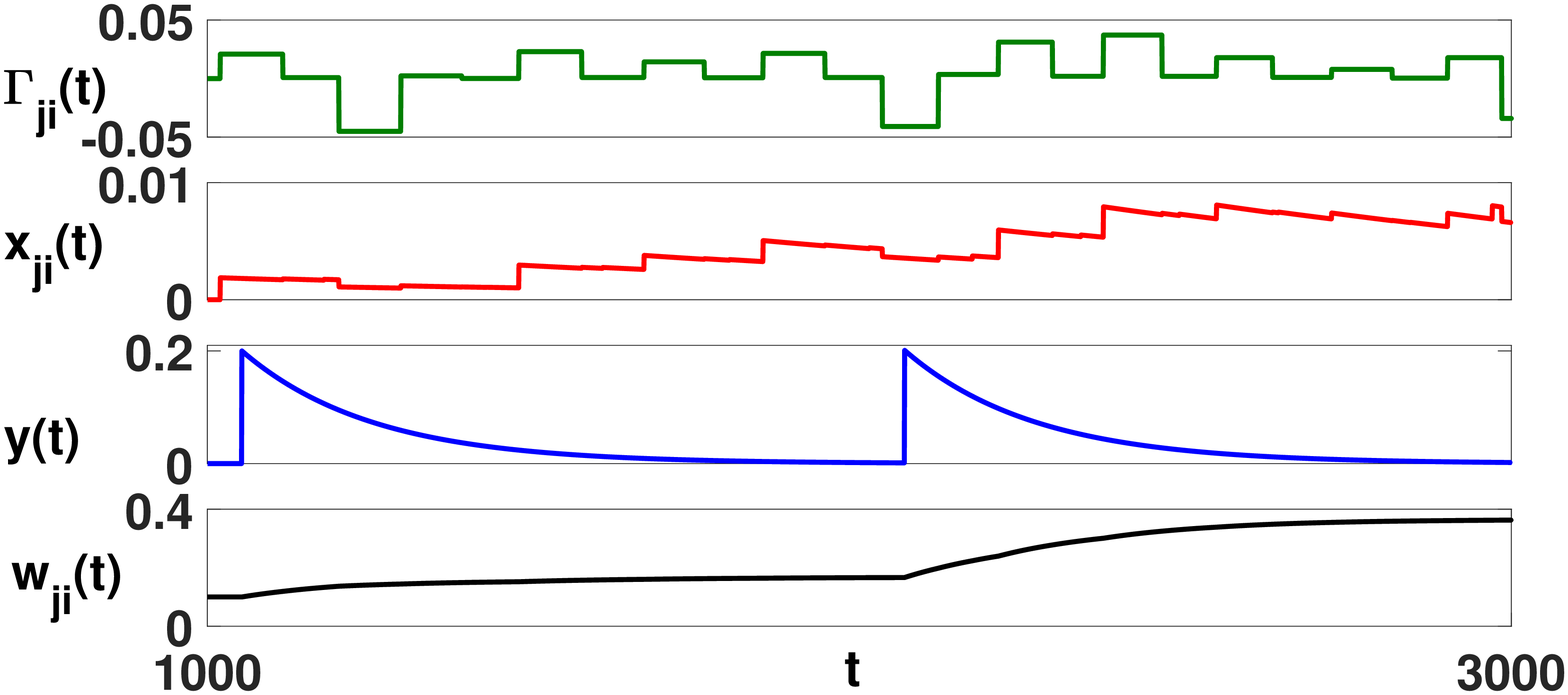}}
\end{center}
\caption{\small Illustration of reward-modulated STDP rule to potentiate a synapse: Temporal order of pre- and post-synaptic spikes determines the value of $\Gamma_{ji}(t)$ (green) and subsequently changes the synaptic tag $x_{ji}(t)$ (red). While the global reward $y(t)$ (blue) is released with a small delay after a correct network's response, if $x_{ji}(t)$ has a relatively large positive value, the synaptic strength $w_{ji}(t)$ (black) increases.} \label{fig1}
\end{figure}

To include the transmission time for a causal relation between
pre- and post-synaptic firing, we employ a temporally shifted STDP
window for which the boundary separating potentiation and
depression does not occur for simultaneous pre- and post-synaptic
spikes, but rather for spikes separated by a small time shift
\cite{Babadi2010}. We set the value of this shift equal with the
actual axonal delay for each synapse. This rule retrieves the
conventional STDP rule when $\tau=0$. This temporal shift causes
synchronous or nearly-synchronous pre- and post-synaptic spikes to
induce long-term depression, which leads to intrinsic stability in
the network \cite{Babadi2010,MVT2018}. Also to keep the
excitation-inhibition ratio balanced during the simulations, the
strength of each inhibitory synapse is set to be $w_{ji} (t)=r
\bar{w}(t)$ while $\bar{w}(t)$ is the average strength of all
excitatory synapses at time $t$.

\textbf{Neural synchronization-} We integrate the dynamical
equations using fourth-order Runge-Kutta method with a time-step
$h=0.1$ ms and obtain $v_i(t)$ and the spike-times of all neurons.
We assign a phase $\phi_i(t)=2\pi\frac{t-t_i^m}{t_i^{m+1}-t_i^m}$
to each neuron between each pairs of successive spikes. $t_i^m$ is
the time that neuron $i$ emits its $m^{th}$ spike. Next we
evaluate a time-dependent order parameter
\cite{KM2019,KM2018,KM2020}:
\begin{equation}\label{sequ7}
S(t)=\frac{2}{N(N-1)}\sum_{i\neq j}cos^2\Big{(}\frac{\phi_i(t)-\phi_j(t)}{2}\Big{)}
\end{equation}
\\
This order parameter measures the collective phase synchronization
at time $t$. $S(t)$ is bounded between 0.5 and 1. If neurons spike
out-of-phase, then $S(t){\simeq}0.5$, when they spike completely
in-phase $S(t){\simeq}1$ and for states with partial synchrony
$0.5<S(t)<1$. The global order parameter $S^*$ is the
long-time-average of $S(t)$ at the stationary state after the
influence of STDP ($S^*=\langle S(t)\rangle_t$).

\textbf{Learning and performance-} Reinforcement of specific
firing patterns is possible by this cognitive system.
In this paper, we reinforce the network to produce an appropriate
response to a stimulus. We choose $n$ random nonoverlapping group
of $5$ neurons, called $S_1, \dots , S_n$ which represent the
input stimulus to the network, and $n$ other random nonoverlapping
group of $5$ neurons, called $R_1, \dots ,R_n$ which give rise to
$n$ distinct responses of the network. $R_i$ is assumed to be the
correct response to stimulus $S_i$ ($i=1,\dots ,n$). Larger $n$
corresponds to more complicated task learned by the network. A
series of simulations carried out with $n$ pairs of
stimulus-response denoted as n-SR simulated experiment and we
consider the cases $n=2,3,4,5$. Our simulations consist of trials
separated by one second. At the onset of each trial the stimulus
is delivered to the network by injection of a strong 2-ms pulse of
current into the neurons in $S_i$. The sequence of $S_i$'s are
delivered to the network randomly. The coincident firing of
neurons in this group typically induces a few spikes in the other
neurons in the network. During a 20-ms time window beginning
$\tau$ ms after the stimulation, we count the number of spikes
fired by neurons in all $R$ groups. We say the network has
exhibited response $R_i$  if $R_i$ has fired the largest number of
spikes among $R$ groups. If in a trial after delivering the
stimulus $S_i$ to the network, $R_i$ and another group $R_j$
simultaneously fire the largest number of spikes, we do not accept
$R_i$ as a correct response and label the case as \textit{no
response}. One might think of $R$ groups as projecting, for
example, to different motor areas in brain. To produce a
noticeable movement, one group has to fire more spikes than the
other groups \cite{IzhikevichDA}. After each stimulation, we
monitor the response of the circuit. If the circuit has exhibited
the desired response, then we deliver a reward in the form of the
increase of extracellular dopamine with a random delay in range
10-50 ms (See Eq.(7)). This delay is included because in
biological neural networks reward typically comes with some delay
after reward-triggering actions. The network performance is
evaluated as \cite{Skorheim2014}:
\begin{equation}
p(T)= p(T-1)(1-\lambda)+ \lambda q
\end{equation}
\\
where, $p(T)$ is the performance in $T^{th}$ trial, $\lambda$ is a
parameter that controls the size of fluctuations of $p(T)$. Although the stationery state amount of $p(T)$ is independent of $\lambda$,
it preferably should be a small value. We set $\lambda=0.002$, $q$ is
a binary variable which is equal to 1 if the network response is
correct and is 0 otherwise.

We note that the intricate details of the model along with the
need to obtain long-time dynamics of the system, limit our
computational abilities. We have therefore performed simulations
for $100<N<400$. We however note that our general results and
conclusions are independent of the system size and will therefore
present results for $N=100,$ and/or $N=200$. A Fortran code is
available upon request.

\section{Results}
First, we establish the occurrence of a synchronization transition
in this cognitive system. The amount of synchronization in neural
networks with static synapses can be controlled upon changing the
average synaptic strength \cite{KM2018,KM2020}. In the present
study synaptic strength is an autonomous variable which is
modified by the internal dynamics of the network through the RL
process. Thus, one cannot control the synaptic strength to tune
the amount of synchronization. However, we have previously shown
that in a similar adaptive network of spiking Izhikevich neurons,
synchronization is a self-organized emergent property which
depends instead on the average time-delay $\tau$ \cite{KM2019}. In
particular, spike-timing dependent plasticity along with time
delay provides an underlying mechanism where the causal effect
(synapse) between strongly correlated neurons gets suppressed,
while slightly correlated activity will be potentiated. Motivated
by this insight, we monitor the amount of synchronization $S^*$
for different values of $\tau$ in the cognitive system and find
that increasing $\tau$ leads to a continuous phase transition from
strong synchronization ($S^* \simeq 0.9$) to asynchronous neural
oscillations ($S^* \simeq 0.5$). Synchronization diagrams of the
circuits with $N=100$ and $N=200$ in different n-SR experiments
are shown in Figs.2(a) and (2b), respectively. It is observed that the transition point $\tau_c$ depends on the system size and
moves toward smaller $\tau$ values by increasing $N$.   We note
that $\tau_c=31$ ms for $N=100$ and $\tau_c=26$ ms for $N=200$ in
all considered n-SR experiments. Increasing the system size
further ($N=400$) reduces the transition point by a lesser amount,
however, computational limits ($N=100,200,400$) do not allow us to
obtain enough point to extrapolate to the thermodynamic limit.

\begin{figure}[!t]
\begin{center}
\subfigure{\includegraphics[width=0.50\textwidth,height=0.24\textwidth]{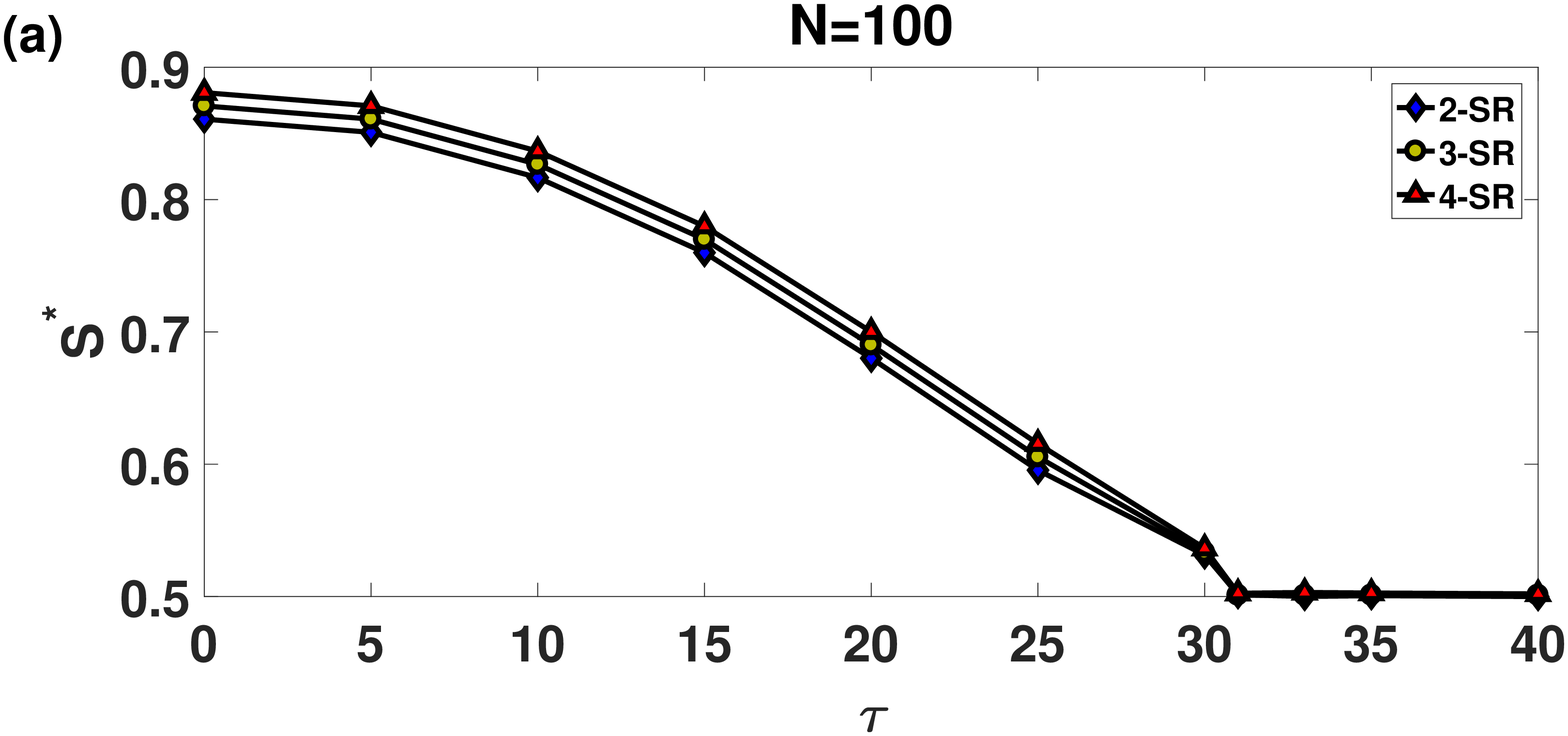}\label{fig2a}}
\subfigure{\includegraphics[width=0.50\textwidth,height=0.24\textwidth]{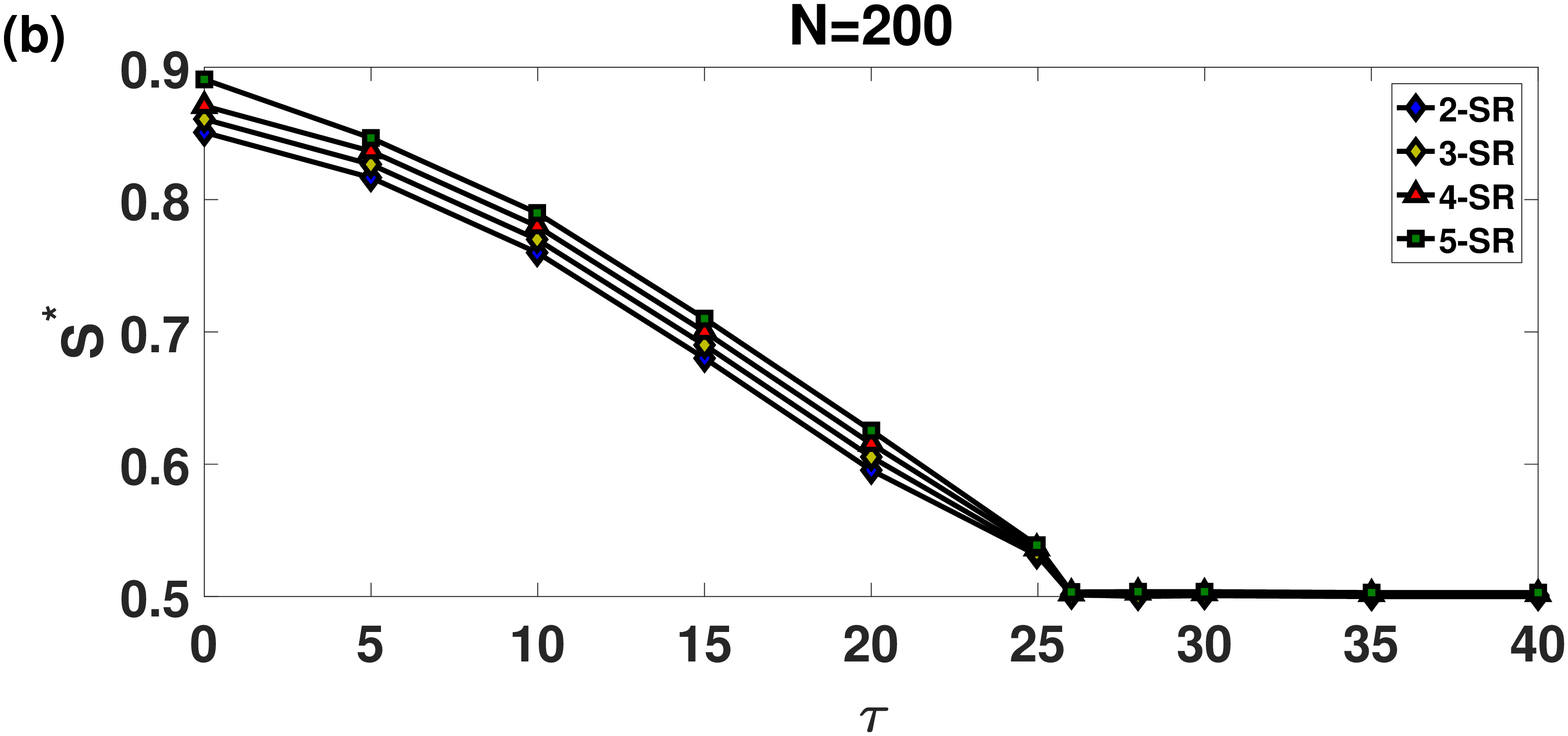}\label{fig2b}}
\end{center}
\caption{\small Synchronization plots
($S^*$ vs. $\tau$) for different
n-SR simulated experiments for the system size (a) $N=100$ and (b)
$N=200$. The dopaminemodulated STDP leads to the emergence of
synchronization transition with a sharp, size-dependent transition
point. $\tau_c(N=100)=31$ ms and $\tau_c(N=200)=26$ ms.}
\label{fig2}
\end{figure}

Next, we show that reinforcement of specific firing patterns is
possible by this cognitive system. Specially, reward-based
reinforcement of the network produces an appropriate response to a
stimulus. Figs.3(a-d) show the performance of the system $p(T)$
vs. trail number $T$ for different $\tau$'s in each n-SR simulated
experiment for $N=200$. Over time, $p(T)$ increases quickly as the
system learns to respond correctly to each stimulus as it reaches
a stationary state after nearly 2000 trials. The stationary value
of $p(T)$ depends on $\tau$. It is seen from Figs.3 the
$p(T)$ vs. $T$  plots for
$\tau=\tau_m=25$ ms (red curve) stands above the other curves for
all n-SR simulated experiments. This is an important indication of
the possible relation between synchronization and learning in the
system. Clearly, the emergence of spontaneous synchronization is
related to learning and its optimization, i.e. performance.

\begin{figure}[!htbp]
\begin{center}
\subfigure{\includegraphics[width=0.50\textwidth,height=0.25\textwidth]{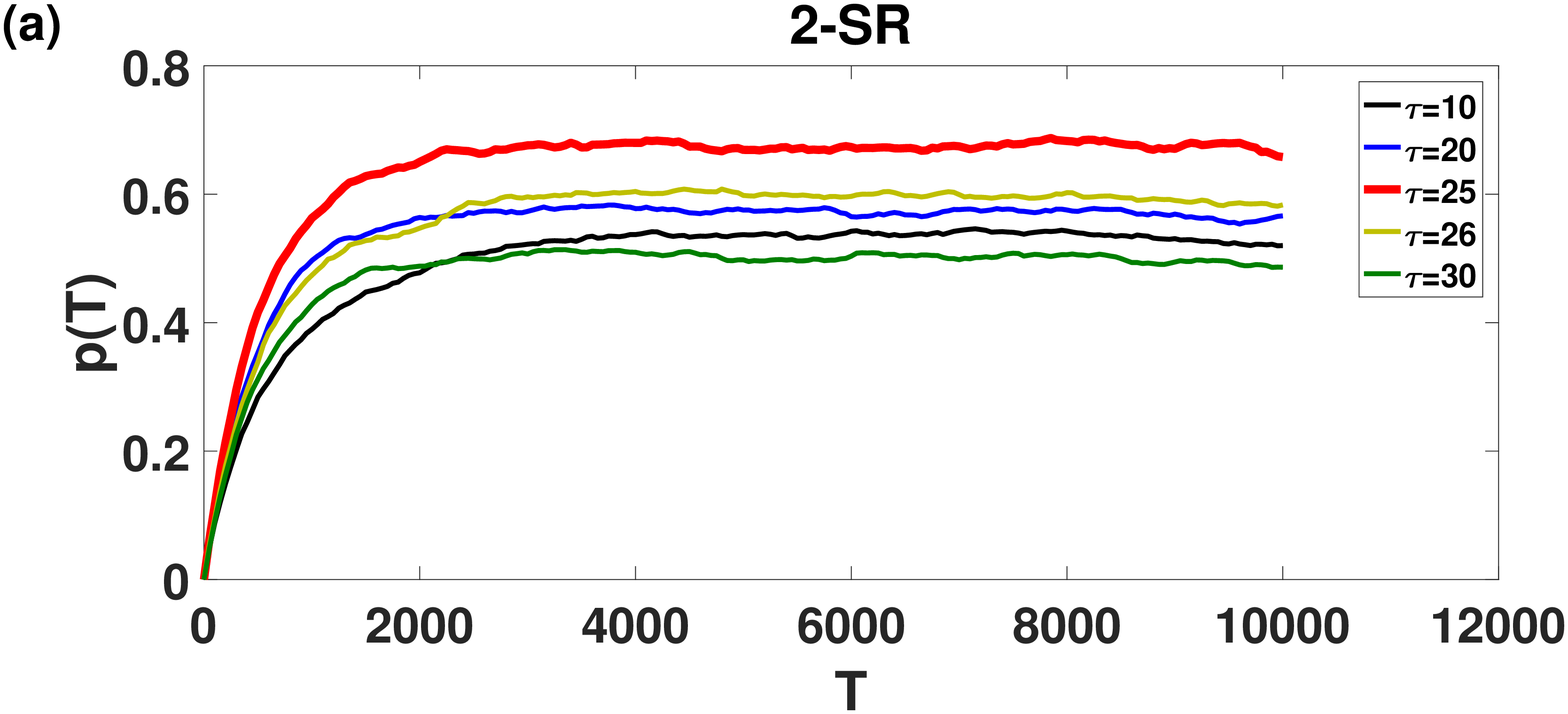}\label{fig3a}}
\subfigure{\includegraphics[width=0.50\textwidth,height=0.25\textwidth]{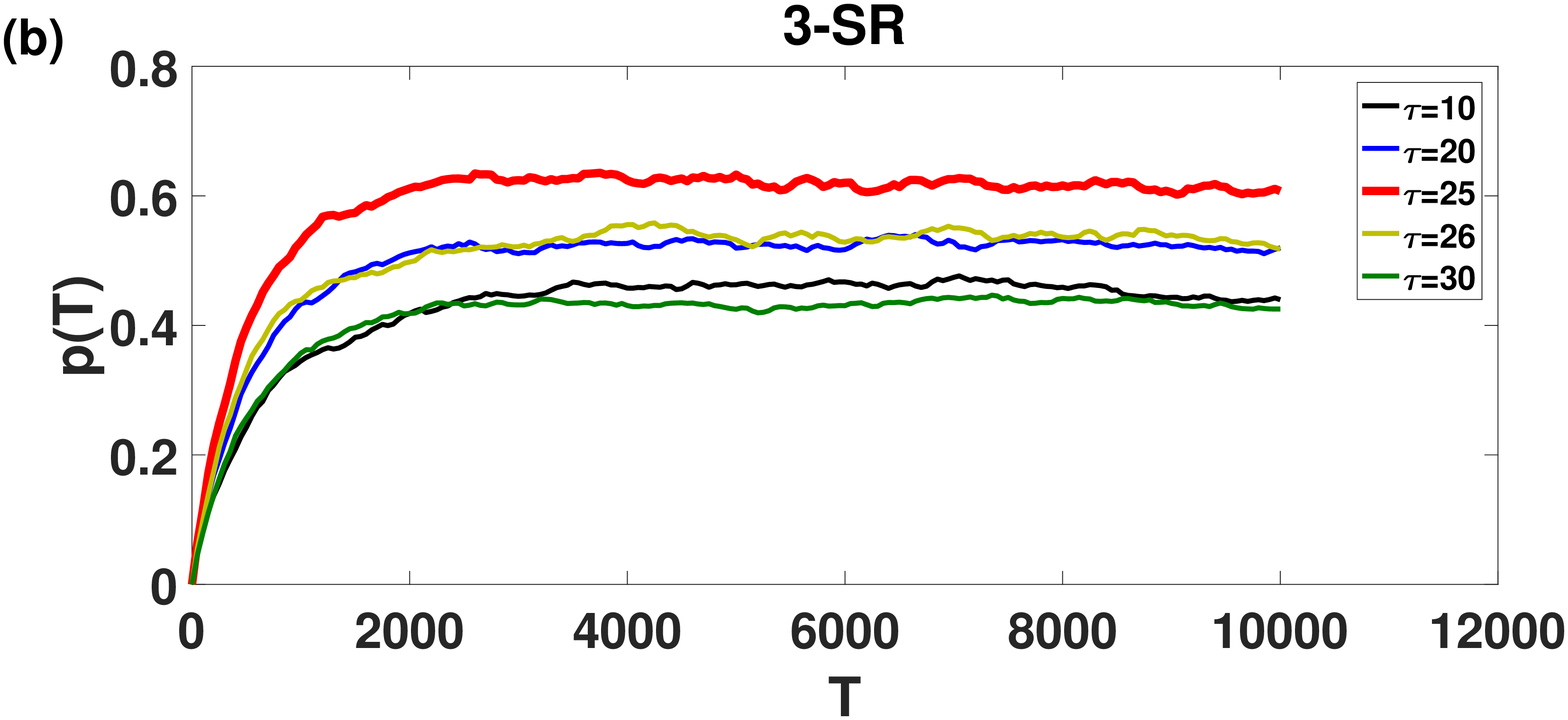}\label{fig3b}}
\subfigure{\includegraphics[width=0.50\textwidth,height=0.25\textwidth]{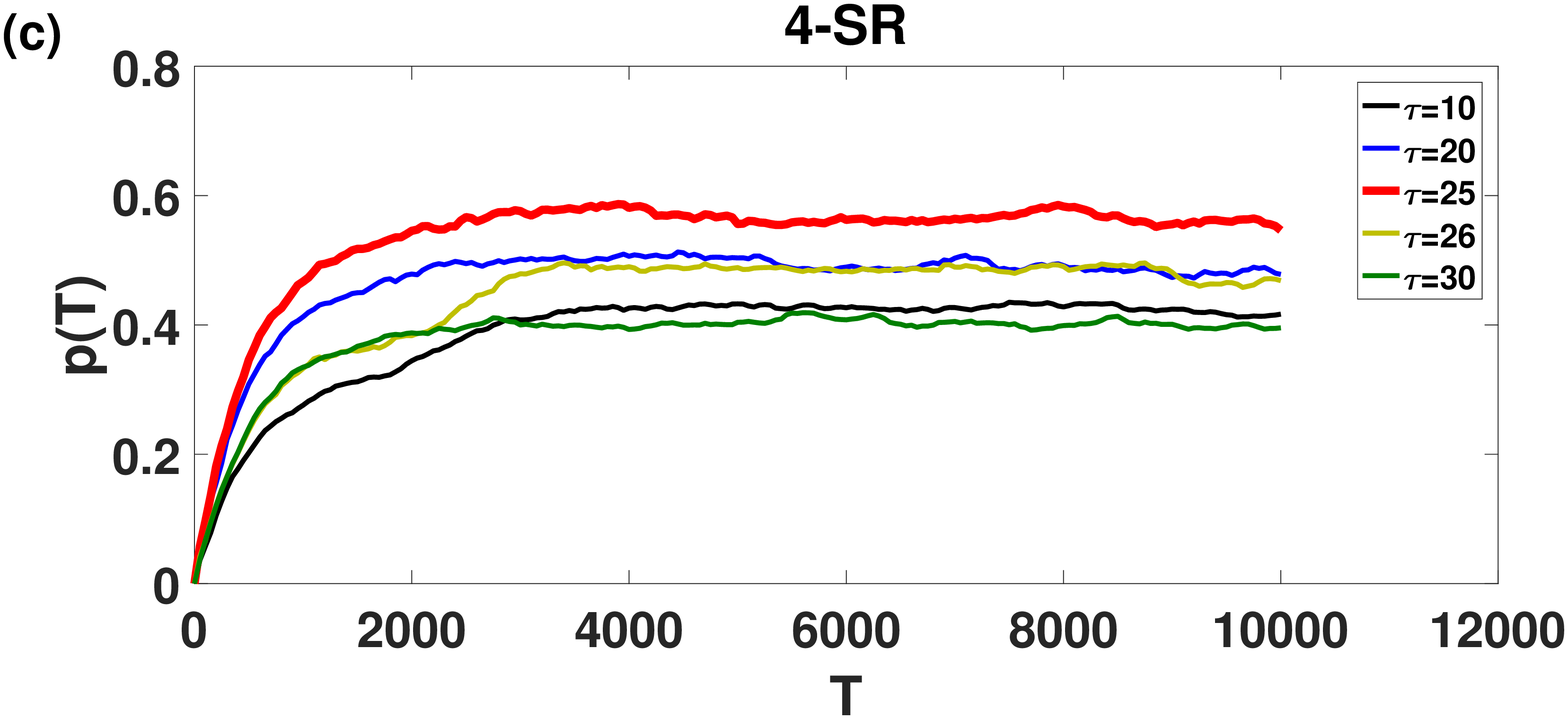}\label{fig3c}}
\subfigure{\includegraphics[width=0.50\textwidth,height=0.25\textwidth]{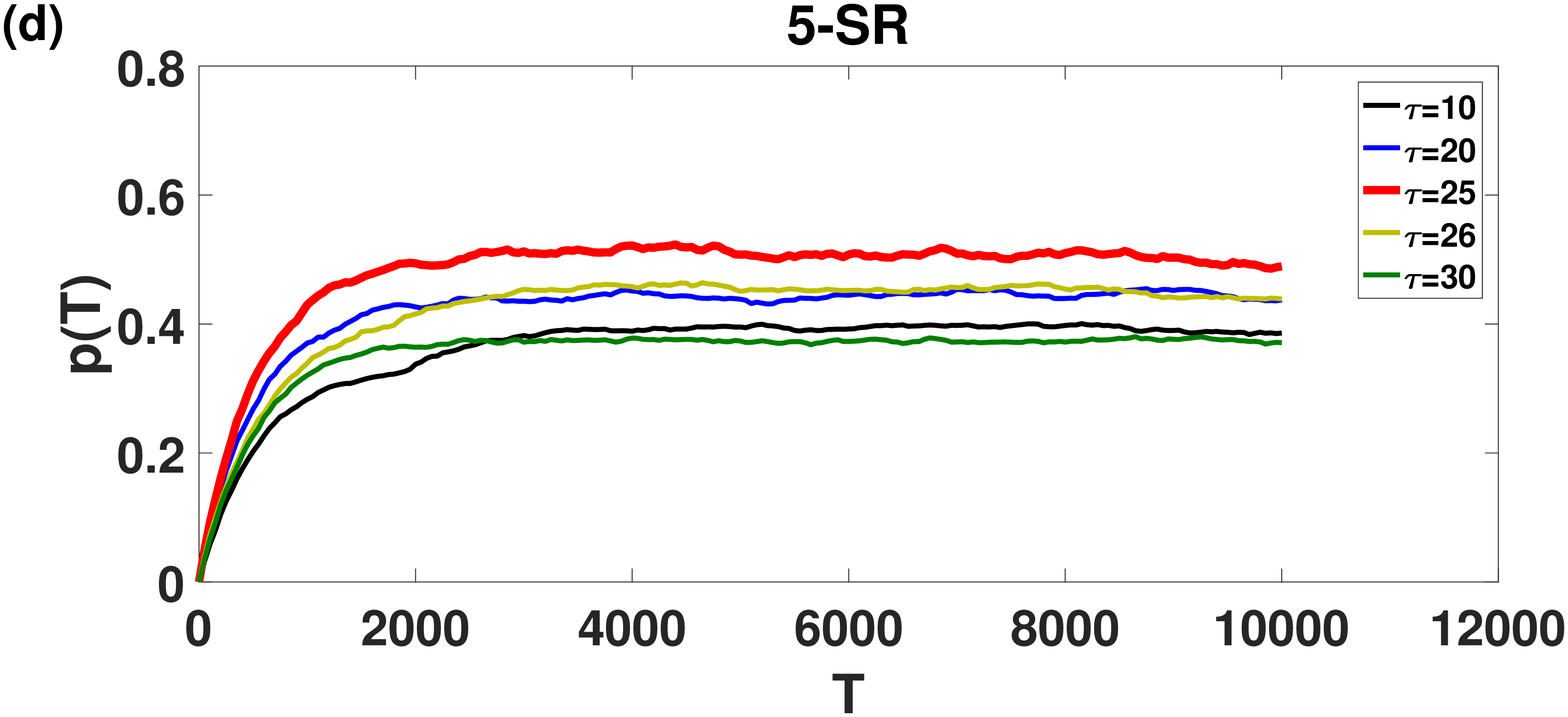}\label{fig3d}}
\end{center}
\caption{\small Dynamics of network performance over time for the
system size $N=200$and for different $\tau$ values in n-SR
simulated experiments with $n=2,\dots ,5$. Time is shown in units
of $T$. To obtain each \textcolor[rgb]{0.00,0.00,1.00}{$p(T)$ vs.
$T$} plot, we carried out an ensemble of 20 simulations with
different stochasticity conditions including network realization.
We coarse-grained the raw data on $T$ axis for each simulation and
next averaged over the ensemble.} \label{fig3}
\end{figure}

It is evident from Figs.2 and 3 that variation of $\tau$ changes
the state of synchronization in the system and subsequently
influences the network's performance. Thus, performance depends on
synchronization. Now, one may ask {\lq\lq}what amount of
synchronization optimizes the performance of the system for a
given SR task?{\rq\rq} To address this question, we define $p^*$
as the average of $p(T)$ in the stationary state ($p^*=\langle
p(T) \rangle _{T>2000}$). Figs.4(a) and 4(b) illustrate
$p^*$ as a function of $\tau$ for
all n-SR simulated experiments in the network with $N=100$ and
$N=200$, respectively. One observes a significant peak at the
value of $\tau_m$ regardless of the value of $n$ or the size of
the system. Interestingly, the value at which performance is
optimized is not exactly at the critical point, $\tau_c$, where
synchronization begins to emerge, but a slightly supercritical
state where a small amount of order is present in the system
$S^*(\tau_m)\approx0.52)$.  For the record, for $N=100$
$\tau_m=30$ where $\tau_c=31$, and for $N=200$ network,
$\tau_m=25$ where $\tau_c=26$, within our numerical resolution.
Also, note that for $N=100$, $n\leq4$ due to limited network size.

Clearly, if one turns off the mechanism for plasticity, no
learning occurs despite the fact that $\tau$ can change and thus
influence the amount of synchronization in the system.  The
performance of the system in this scenario is a random performance
which tends to decrease as the learning task gets more
complicated, i.e. increasing $n$.  However, for a given $n$ one
can evaluate the performance of the system in order to compare how
plasticity plays a role in the learning performance. This is shown
in Fig.4(c) for $N=200$. In order to better characterize the
relative performance of the system, we define
$\eta(n)=p^*_m(n)/p^*_{un}(n)$, where $p^*_m(n)=p^*(\tau_m)$ and
$p^*_{un}$ is the maximum of the untrained, random performance
depicted in Fig.4(c) for each $n$-task. In Fig.4(d) we show the
results for the relative performance as a function of $n$ for the
two system sizes we have studied. The results indicates that the
relative performance increases with increasing $n$, and for a
given $n$ with increasing system size. Therefore, optimization may
lead to significant improvement in learning of complicated tasks
in large networks. We also note that one can simply define
$\eta(\tau)=p^*(\tau)/p^*_{un}(\tau)$, for various $n$-task
performance (see inset of Fig.4(d)). Incidentally, we note that
the relative performance due to plasticity can increase from
$\eta(\tau=40)\approx1.3$ to $\eta(\tau_m)\approx2.1$, i.e. a
difference between a thirty percent improvement and a more than
hundred percent improvement for a given task by taking advantage
of criticality.

\begin{figure*}[!t]
\begin{center}
\subfigure{\includegraphics[width=0.49\textwidth,height=0.266\textwidth]{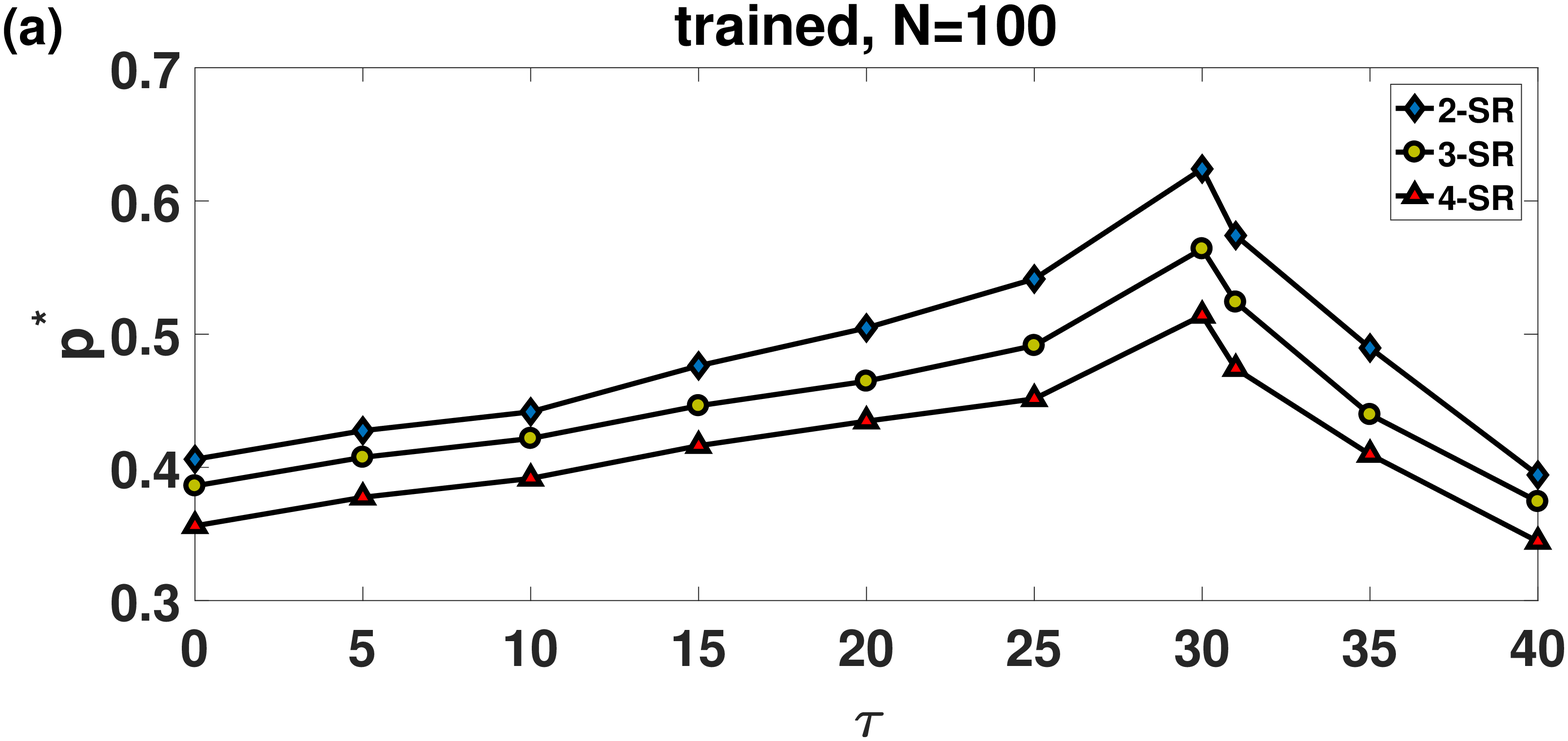}\label{fig4a}}
\subfigure{\includegraphics[width=0.49\textwidth,height=0.266\textwidth]{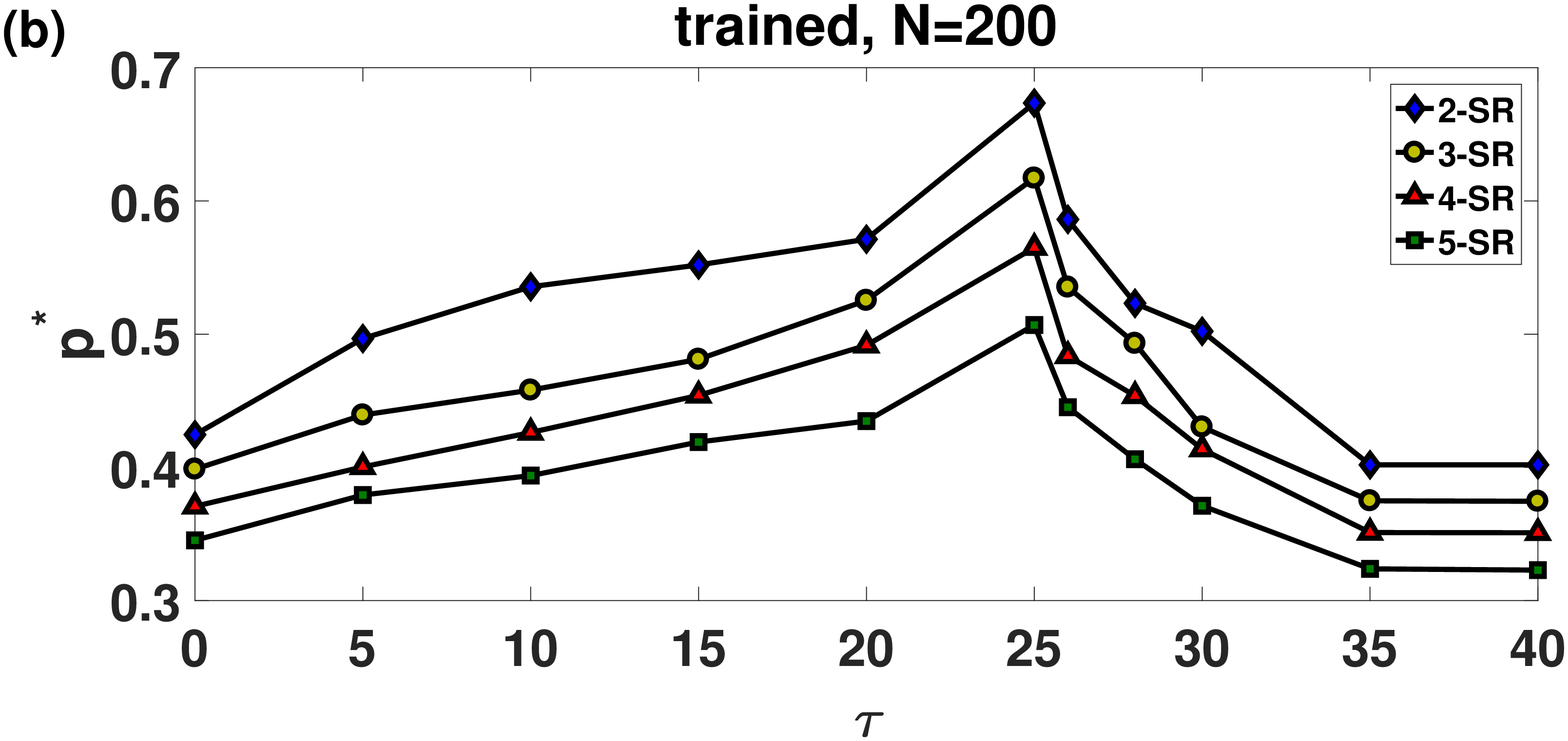}\label{fig4b}}
\subfigure{\includegraphics[width=0.49\textwidth,height=0.266\textwidth]{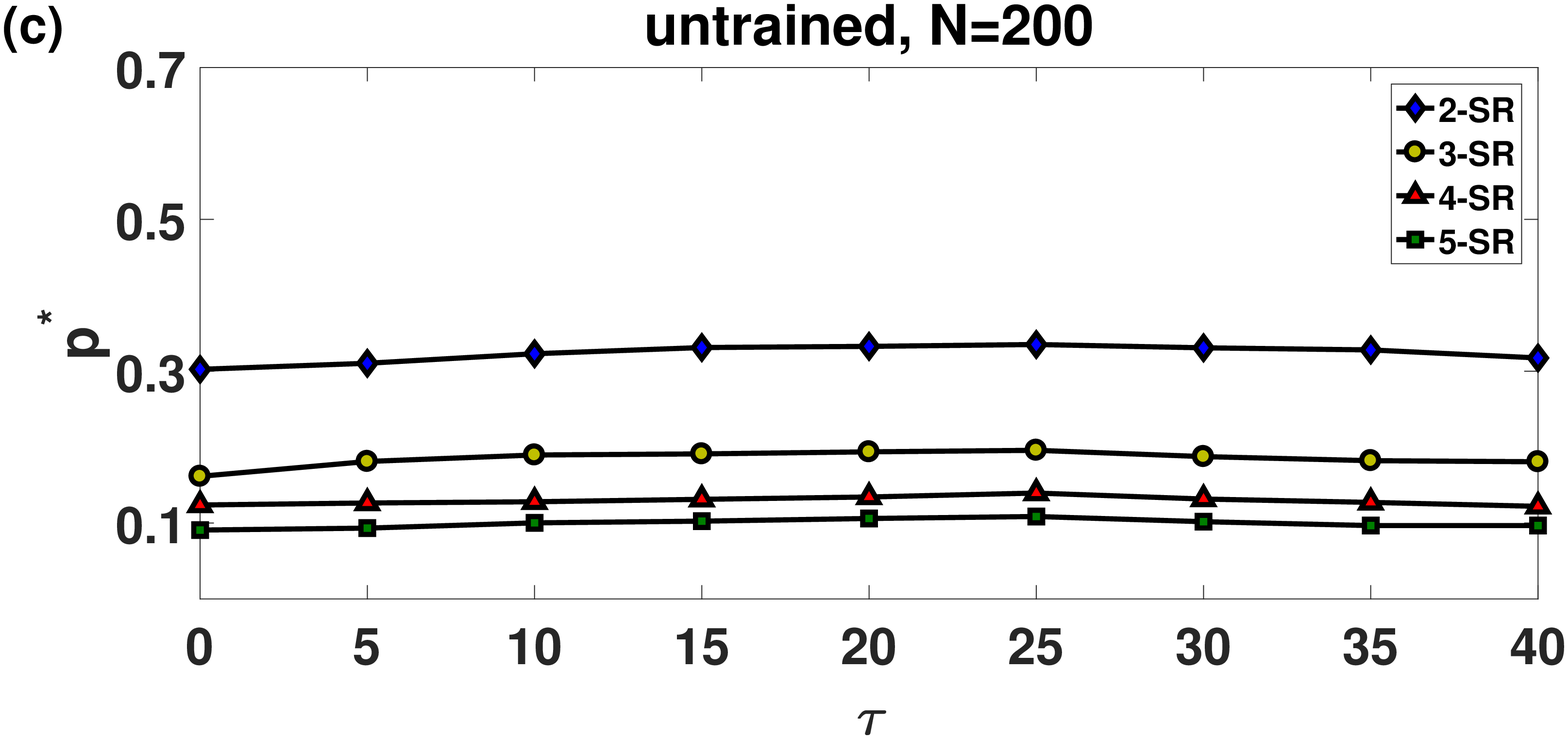}\label{fig4c}}
\subfigure{\includegraphics[width=0.49\textwidth,height=0.266\textwidth]{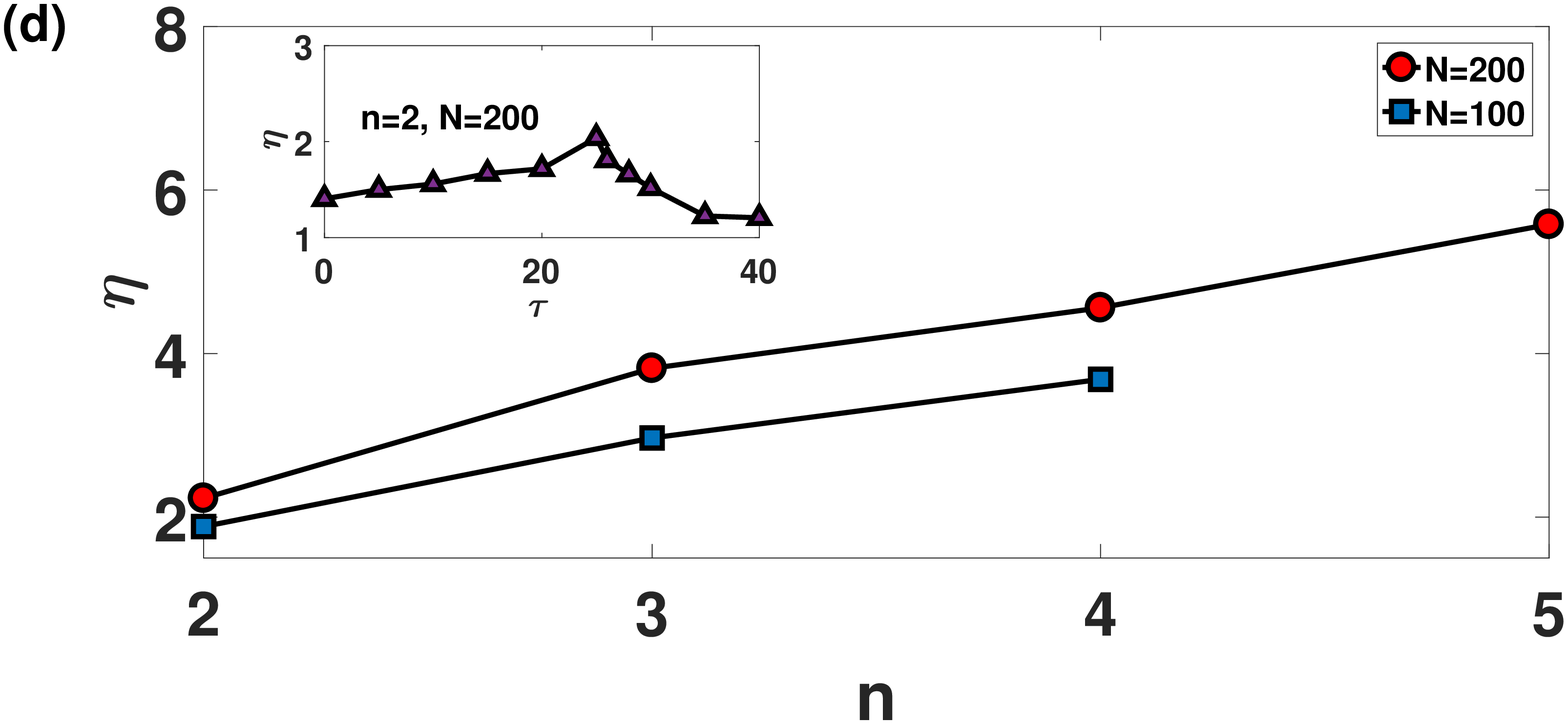}\label{fig4d}}
\end{center}
\caption{\small \textcolor[rgb]{0.00,0.00,1.00}{$p^*$ vs. $\tau$}
in n-SR simulated experiments for: (a) trained system with $N=100$
and $n=2,3,4$. (b) trained system with $N=200$ and $n=2,3,4,5$..
(c) untrained system with $N=200$ and $n=2,3,4,5$. (d)
$\eta$ vs. $n$} plots that show
the relative efficiency of the model improves with increasing $n$
as well as system size $N$. (Inset: The relative efficiency of the
model vs. $\tau$ for the system with $n=2$ and $N=200$.)
\label{fig4}
\end{figure*}

Many recent studies have indicated the functional advantages of
operating near the critical point associated with second order,
critical phase transition. Such studies both include computational
models \cite{Munoz2018} as well as clinical
studies\cite {Zimmern2020}. Our results seem to indicate such
functional advantage in a biologically relevant spiking neural
network.  However, our results also indicate a slightly
supercritical state in a synchronization transition as the optimal
point of reinforcement learning. A standard method to characterize
the critical state is to look at neuronal
avalanches \cite{Beggs2003,Beggs2004,Plenz2014,PRL2019,Friedman2012,Chialvo2013}.
We next propose to study neuronal avalanches in order to better
correlate critical dynamics with reinforcement learning.

Neural avalanches are heterogeneous outbursts of activity
interspersed by brief periods of quiescence
\cite{Beggs2003,Beggs2004,Plenz2014}. To study neural avalanches,
we record the raster plot in the stationary state of the system in
each simulated experiment. Next, we divide it into temporal bins
of length $5$ ms and count the number of spikes in each bin to
extract the time-series of network activity $M$, see Figs.5(a) and
5(b). By monitoring the network activity $M$ we can identify
outbursts of spikes the number of which is associated with the
size $s$, and the lifetime with the duration $d$ of avalanches. An
avalanche begins when $M$ exceeds a threshold $M_{th}$ and ends
when it turns back below that threshold, Fig.5(b). Here, we set
the threshold to be $M_{th}= \overline{M} - \sigma$ while
$\overline{M}$ and $\sigma$ are the mean value and standard
deviation of $M$, respectively. Our consideration shows that
displacement of the threshold in the interval
$[\overline{M}-\sigma, \overline{M}]$ does not influence the
avalanches statistics.

\begin{figure}[!htbp]
\begin{center}
\subfigure{\includegraphics[width=0.50\textwidth,height=0.26\textwidth]{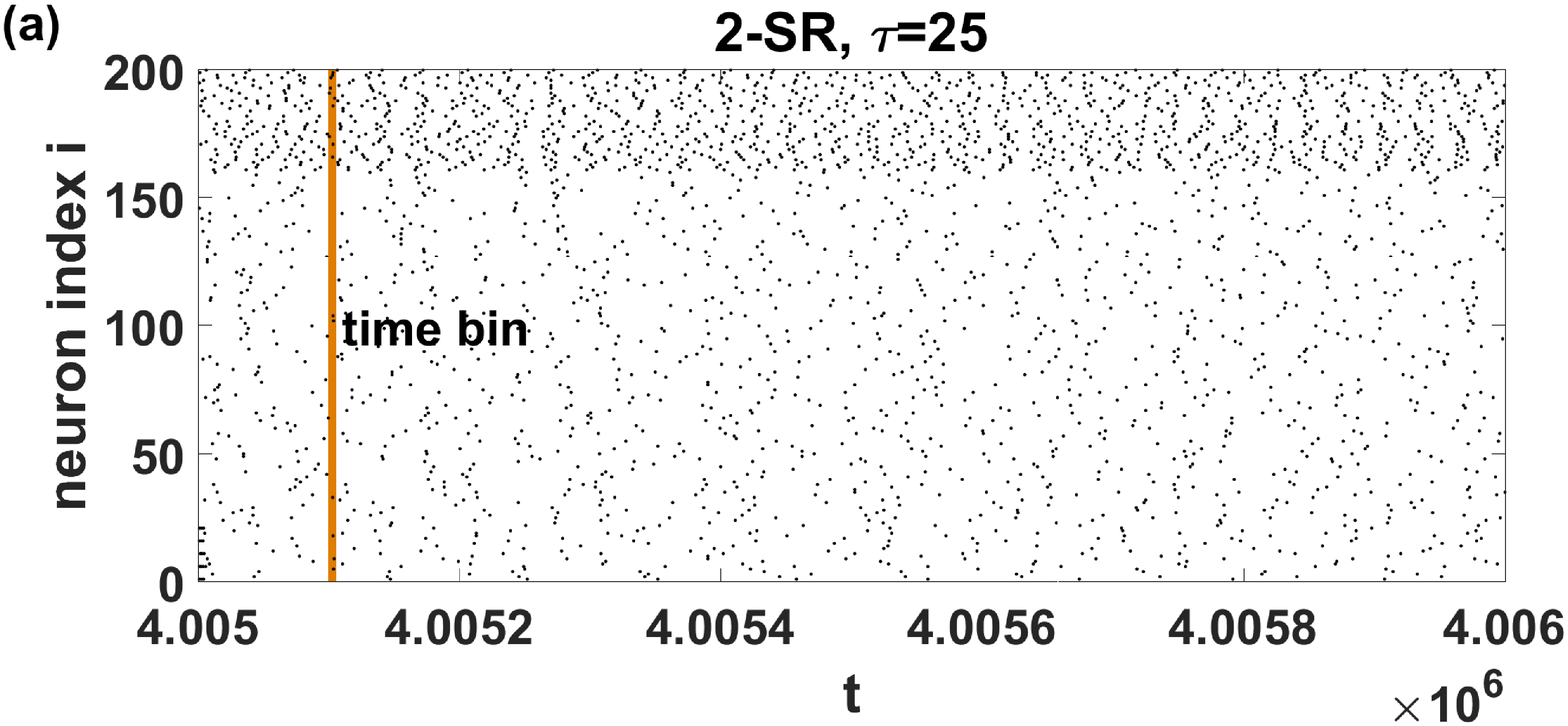}\label{fig5a}}
\subfigure{\includegraphics[width=0.50\textwidth,height=0.26\textwidth]{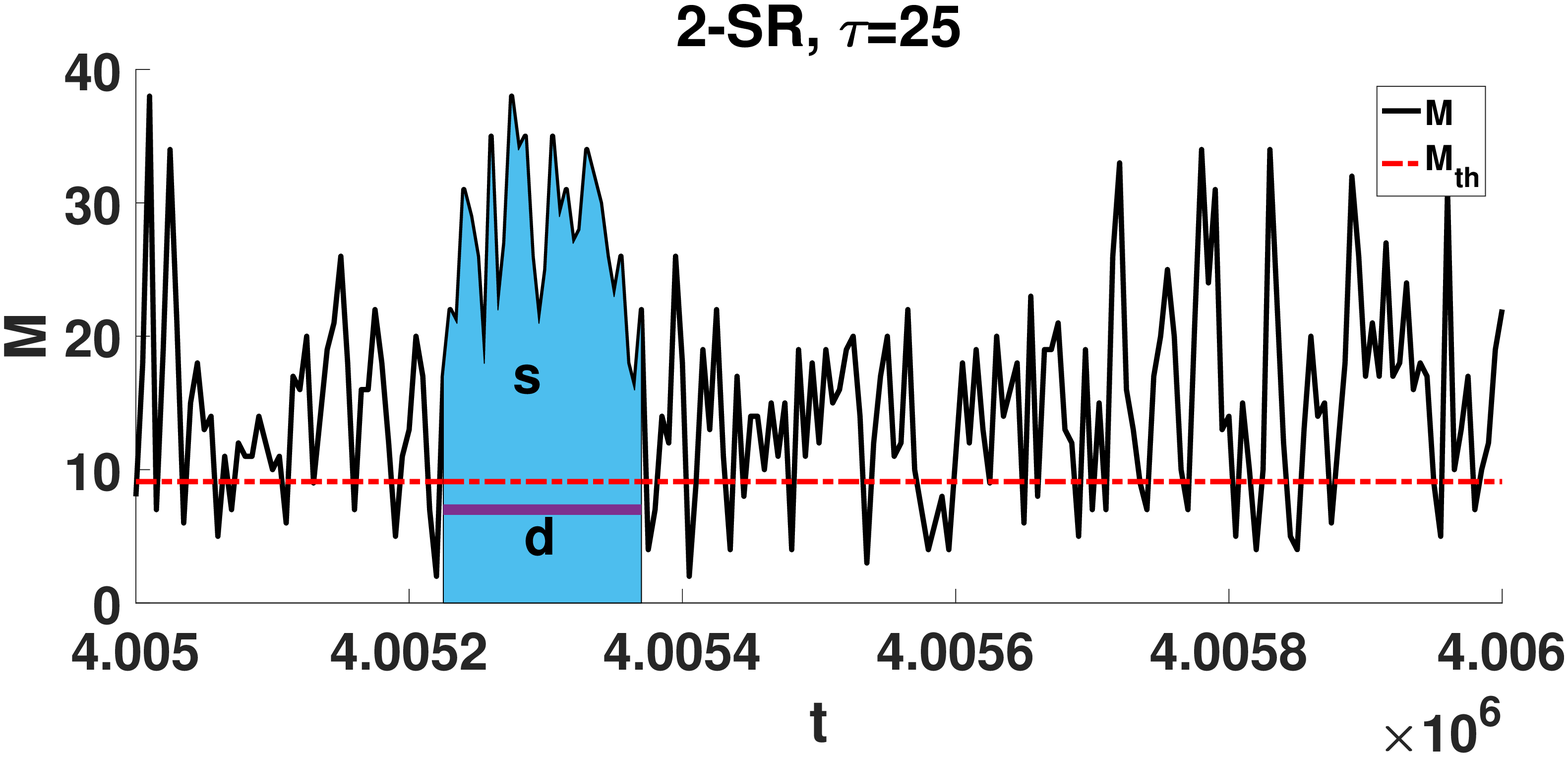}\label{fig5b}}
\subfigure{\includegraphics[width=0.50\textwidth,height=0.26\textwidth]{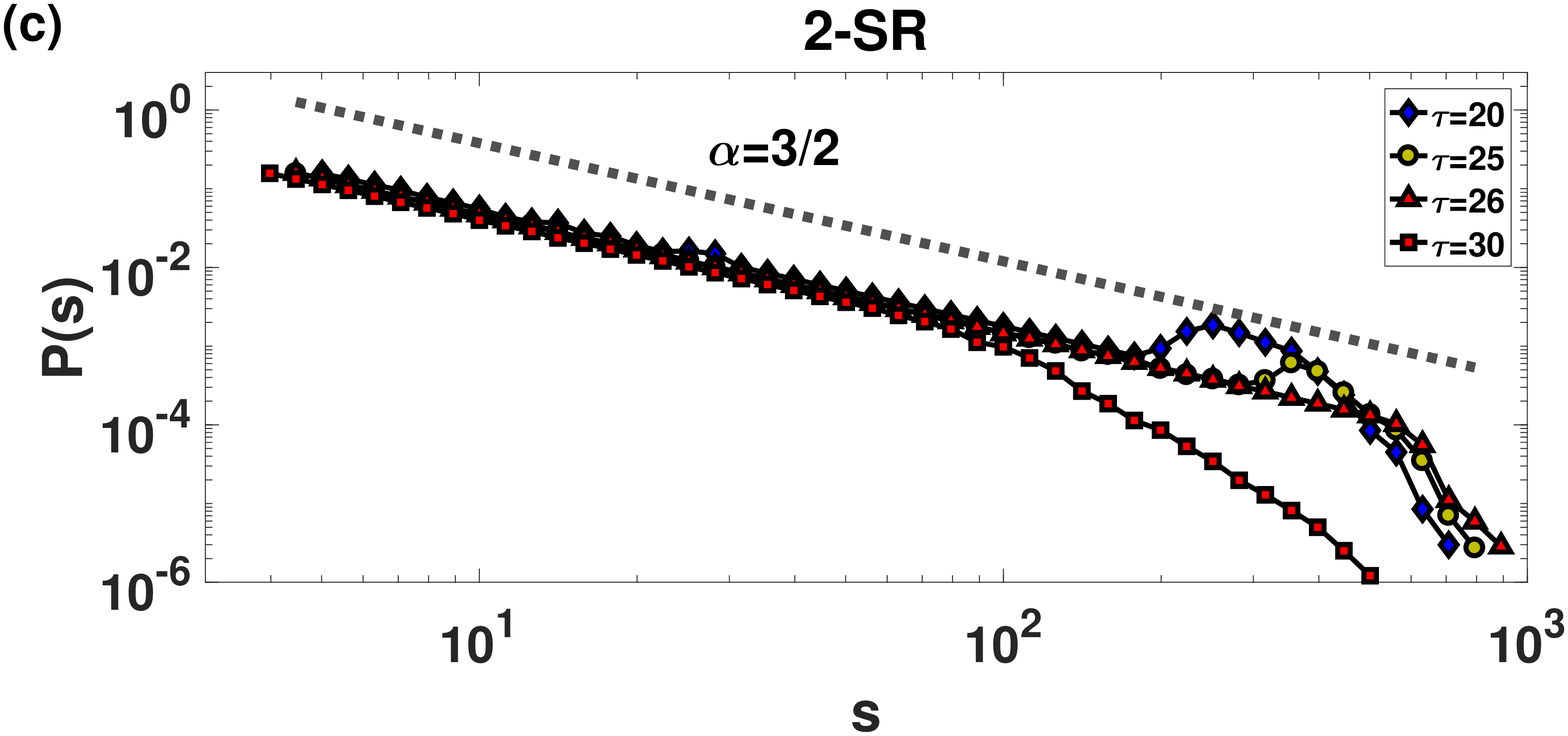}\label{fig5c}}
\subfigure{\includegraphics[width=0.50\textwidth,height=0.26\textwidth]{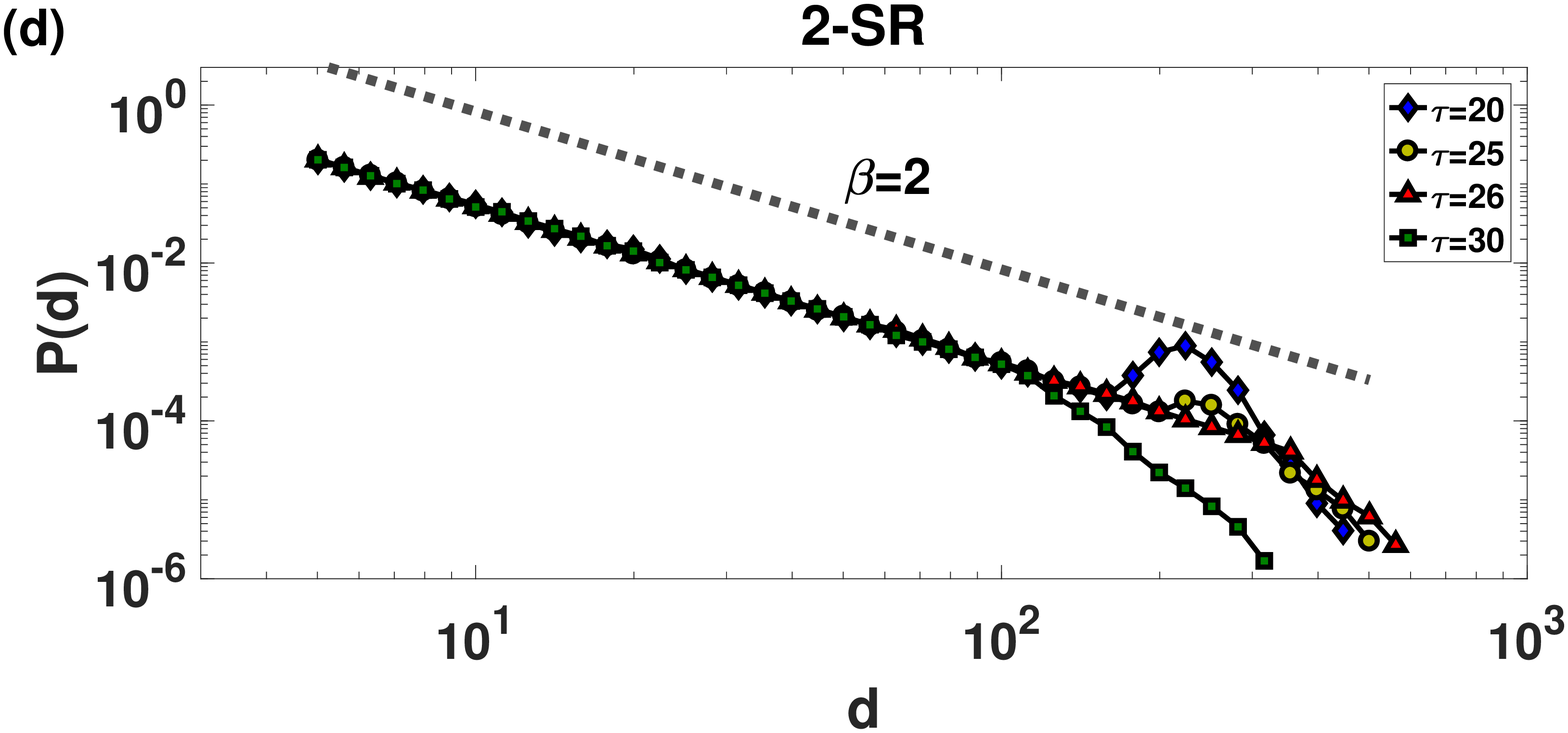}\label{fig5d}}
\end{center}
\caption{\small (a) A 1s segment of the raster plot for the system
that learns 2-SR task. The orange area illustrates a typical time
bin used to extract $M$ time-series.  (b) A typical snapshot of
the network activity $M$ and the threshold $M_{th}$. The blue area
shows the size $s$ of an avalanche and the length of purple line
shows its duration $d$. (c) and (d) Probability distribution
function of avalanches size $P(s)$ and avalanches duration $P(d)$
for the trained networks performing 2-SR task. The network size is
$N=200$. Avalanches are measured only when the performance has
reached a stationary state. The dashed lines are meant to help the
eye.} \label{fig5}
\end{figure}

We consider neural avalanches for different values of $\tau$ and
different n-SR simulated experiments. For any given set of
parameters the network is simulated for a considerably long time,
producing nearly five million avalanches. Probability distribution
of avalanche sizes and avalanche durations for 2-SR simulations of
the network with $N=200$ are illustrated in Figs.5(c) and 5(d),
respectively. $P(s)$ and $P(d)$ for other SR tasks are
qualitatively the same (not shown). The dotted-lines in Figs.5(c)
and 5(d) depict power-law functions $P(s)\sim s^{-\alpha}$ and
$P(d) \sim d^{-\beta}$ with $\alpha=3/2$ and $\beta=2$. These
exponents are associated with mean-field branching process which
are thought to underlay the critical dynamics of such networks \cite{Moosavi2014,Moosavi2015}.
For both size and duration statistics (Figs.5(c) and 5(d)), we
show avalanche statistics for critical ($\tau=\tau_c=26$),
subcritical ($\tau=30$), as well as supercritical ($\tau=20$)
states. The critical state is associate with a power-law limited
by the finite size of the network at large events. Subcritical
state is associated with a cutoff and thus decaying large events,
while the supercritical state is associated with a significant
``bump" at large events. We have also plotted the avalanche
statistics for the corresponding value of time delay which
optimizes RL performance, i.e. $\tau=\tau_m$. As expected, it
displays a very similar statistics to the critical state with a
slight bump towards the large events. We therefore conclude that
being near the critical point offers the cognitive
system functional advantages which are not present when operation
away from criticality. The fact that such optimization occurs in a
slightly supercritical state could be related to the occurrence of
more frequent large avalanches (i.e. small bump) which may play a
role in correlated activity at large distances which facilitate
pathways between S and R neurons, while still keeping the usual
advantages associated with the actual critical point. We finally
note that similar changes in the profile of avalanche statistics
due to task performance have been observed in various experimental
studies \cite{Yu2017,Shew2015,Arviv2015}. However, a clear method
to measure ``the distance" to criticality is still lacking.

\section{Concluding Remarks}
In this work we have proposed to study a biologically relevant
dynamical system which learns to exhibit specific responses to
stimuli via reinforcement learning. The Izhikevich spiking
neuronal network is supplemented by time-delayed chemical synapses
as well as dopamine modulated, time-delayed, spike-timing
dependent plasticity. We observe that the system can show various
amount of synchronization depending on the average time delay,
with a continuous phase
transition to an asynchronous state. Learning performance is then
evaluated in the steady state and is shown to depend on the amount
of synchronization in the system. Interestingly, it is found that
regardless of the task complication ($n$) or network size studied
($N$), optimized performance for learning occurs near the critical
transition point. In order to better understand such behavior we
studied avalanche statistics as a way to probe the critical
behavior of the system.  We observe subcritical and supercritical
avalanche statistics away from the critical point which itself
exhibits clear power-law behavior with experimentally relevant
exponents. We observed that the optimal learning state corresponds
to a near critical state with a slight bump in the tail of the
statistics, indicating the significance of more probable large
events which may help correlate the distinct areas of the network
associated with stimulus and response. This type of correlated
activity provides more frequent long-range pathways which
facilitate efficient learning.

A few points are of note. First, the reinforcement learning model
which we have used \cite{IzhikevichDA}, is based on reward
modulation of a purely Hebbian STDP rule with a reward signal that
is always positive. Another possible approach to implement
reinforcement learning through reward-modulated STDP is given by
Florian \cite{Florian2007}, with Hebbian STDP when the reward is
positive, and anti-Hebbian STDP when the reward is negative. A
neural network trained using a simplified version of Florian
algorithm has shown to be a competitive alternative to the
optimized state-of-the-art models in the field of machine learning
\cite{Chevtchenko2021}.

Second, in comparison to the
typical models employed in physics, the model considered in the
current study is a detailed model which is quite complicated, in
the sense that it is high dimensional and possesses many
parameters. However, several parameters including the parameters
of the Izhikevich neurons (Table 1) which are set to produce
regular spiking pattern of firing as well as the parameters of
synaptic currents (Table 1) and dopamine signaling (Eqs.(6) and
(7)) that are set to be compatible with empirical findings, are
fixed in all simulations. We carried out intensive computer
simulations to examine the effect of other parameters on the
system's behavior. We found that the above results are robust upon
changing STDP parameters (Table 1), and upon changing the average
connectivity of the initial ER network. We assessed the
statistical properties of the networks before and after learning
and observed that the trained networks still remain as random (ER)
networks with Poisson degree distribution functions. The average
connectivity, $p$, however typically decreases as a result of
pruning caused by learning. Additionally, we observed that the
above results are preserved while we start with a complete network
rather than an ER network.

Third, critical behavior and its associated
benefits in neural networks does not necessarily come by operating
exactly at the critical point. Various mechanisms exist which can
extend the ``critical regime". Griffith's phase is the most
well-known such mechanism which has structural origins
\cite{Munoz2010,Buendia2021}. There is also dynamical mechanisms
such as refractoriness in neuronal oscillations which can extend
the critical regime \cite{Moosavi2017}. However, what we have
observed is that a slight deviation from standard critical scaling
regime leads to a more efficient learning.

Lastly, further studies are needed to better understand the
relevance of synchronization and criticality, with reinforcement
learning. Specifically, with regards to the functional advantages
associated with the critical dynamics of neuronal networks, and
recent intensive studies in machine learning, one would expect
that insights from criticality literature may find important
ramifications in the field of machine learning. However,
neuromorphic networks may provide an ideal system to
experimentally study oscillations, learning and avalanches in a
controlled manner \cite{Hochstetter2021}.

\section{Acknowledgements}
Support from Shiraz University research council is acknowledged.

\bibliographystyle{apsrev}
%\bibliography{xbib}

\end{document}